\documentclass[aps,prd,onecolumn,showpacs,amssymb]{revtex4}
\usepackage{graphicx,bm,color}
\usepackage{amsmath}
\usepackage{amssymb}
\usepackage{amsfonts}
\usepackage{epsfig}
\newcommand{\be}{\begin{equation}}
\newcommand{\ee}{\end{equation}}
\newcommand{\bea}{\begin{eqnarray}}
\newcommand{\eea}{\end{eqnarray}}
\newcommand{\beaa}{\begin{eqnarray*}}
\newcommand{\eeaa}{\end{eqnarray*}}

\newcommand{\nn}{\nonumber \\}

\allowdisplaybreaks[4]

\begin{document}

\title{On the stability of Einstein static universe at background level in massive bigravity  }
\author{ M. Mousavi}\email{ mousavi@azaruniv.edu}
\author{F. Darabi}\email{ f.darabi@azaruniv.edu}

\affiliation{Department of Physics, Azarbaijan Shahid Madani University, Tabriz, 53714-161 Iran}

\begin{abstract}
We study the static cosmological solutions and their stability at background level in the framework of massive bigravity theory with Friedmann-Robertson-Walker (FRW) metrics. By the modification proposed in the cosmological equations subject to a perfect fluid we obtain new solutions  interpreted as the Einstein static universe. It turns out that the non-vanishing size of initial scale factor of Einstein static universe depends on the non-vanishing three-dimensional spatial curvature of FRW metrics and also the graviton's mass. By dynamical system  approach and numerical analysis, we find that the extracted solutions for closed and open universes can be stable { for some viable ranges of equation of state parameter, viable values of  fraction of two scale factors, and viable values of graviton's mass obeying the hierarchy $m << M_Pl$ which is more cosmologically motivated.}
\end{abstract}


\maketitle

\section{Introduction \label{Sec1}}
General relativity predicts that at the beginning of the universe all mass, energy, and spacetime  were compressed to an infinitely dense point at  the Planck epoch, called
initial singularity. Quantum mechanics becomes a significant factor at the Planck epoch and is hoped to help us in avoiding the classical singularity arising from application of general
relativity to the system of universe. In response to the incompetence of general relativity in the study of initial state of the universe,  alternative theories for the beginning of the universe have been proposed which are basically based on the
 application of quantum theory on the system of universe.
The quantum cosmology is one of those  general frameworks which has been
widely used to resolve the initial singularity problem \cite{QC}. Furthermore, the string/M-theory, the pre-big bang \cite{1} and ekpyrotic/cyclic \cite{2} scenarios have been proposed to  resolve this problem. One of the recent
approaches, so called ``emergent universe'', to avoid the big bang singularity was proposed by Ellis {\it et al} in the framework
of general relativity \cite{3,4}. The emergent universe is a scenario in which the space curvature is positive, and the universe stays past-eternally in an Einstein static state after which it evolves to an inflationary phase. Thus, this theory is consistent with an inflationary scenario in which the initial singularity is replaced by an initial state, so called Einstein static universe.
Einstein static universe is the exact solution of Einstein equations equipped by closed Friedmann-Robertson-Walker metric, a cosmological constant and
perfect fluid matter \cite{6}. In spite of its static feature, it is critically unstable against small perturbations. Hence, in order to avoid  possible
collapse of this static solution toward a singularity, namely in order for the static solution  can play the role to avoid initial singularity we have to investigate
its stability conditions. In this regard, using the modified cosmological equations,  some new static solutions with suitable stability properties
are obtained \cite{13'}.

In the framework of general relativity, the Einstein static scenario suffers from a fine-tuning problem. This problem is alleviated when the cosmological equations of general relativity (GR) are modified within the context of modified theories of gravity. For this reason, the existence of analogous Einstein static solutions in several modified gravity theories and quantum gravity models have been investigated and studied \cite{7,8,9,10,11,12,13,Shab}. It has been found that, depending on the details of the modified gravity theories, the modified cosmological equations result in many new static solutions whose stability properties are substantially different from those of  Einstein static solution of GR. Recently, the stability of Einstein static universe in massive gravity theory has been studied in Ref.\cite{26}. Along this line of activity, we are motivated to study the stability of Einstein static universe in massive bigravity theory. The relevance of this study lies in the fact that contrary to the massive gravity model which has one scale factor, in massive bigravity model we have two scale factors each of which can lead independently to the singularity problem, so it seems that the  massive bigravity model is more concerned than the massive gravity theory, regarding the singularity problem. This may justify the present study of Einstein static universe to avoid the singularity problem in the context of massive bigravity theory. Moreover,  this study is motivated by the possibility that the universe
might have started out in an asymptotically Einstein
static state, well above the quantum gravity scale, in the context of inflationary universe \cite{3}.

A consistent theory describing a massive spin-2 particle was first introduced by Fierz and Pauli in 1939 \cite{15} and developed to a covariant theory by de Rham, Gabadadze and Tolley (dRGT) \cite{16,17} in 2010. The covariant massive gravity model has been proposed in a consistent form in which nonlinear terms have been tuned to eliminate order by order the negative energy states in the spectrum \cite{18}. In the case of massive gravity the graviton mass typically plays an important role over cosmological scales at late times which can lead to the presently observed accelerating phase \cite{27}. Moreover, the theory results in some exotic solutions in which the graviton mass contribution influences the cosmological dynamics at early times.

Massive gravity generically suffers from strong coupling problems and a loss of predictivity at low scales. Indeed, massive gravity  models, depending on the mass scale of the graviton, can have a rather low strong coupling scale which severely restricts the applicability of massive gravity
and also makes it problematic to investigate early universe high energy setup .    The strong coupling scale of bigravity is not known, although some people conjectured that it is the same as in massive gravity. For example, in a recent work \cite{cusin} the inflationary perturbations in bigravity model has been studied
where the energy scale of inflation is typically above the low strong coupling scale of the bigravity theory. Therefore, bigravity has become strongly coupled at
the scale of inflation and hence it is not granted that cosmological
perturbation theory applies. The authors in \cite{cusin} have claimed that since the strong coupling scale is derived in a Minkowski
background, it is not clear whether the strong coupling scale represents an upper limit on the Hubble scale or it is just an upper limit on the energy of the perturbations on a given background.  In another work \cite{Yas}, the authors have argued that, contrary to intuition gained from massive gravity, at energy scales relevant to cosmology the bimetric models can avoid the known strong
coupling issues, namely,
  as long as these models are used for late-time cosmology (or for early times with care), the low strong coupling scale is not a serious
problem. In any case, the problems related to the low strong coupling scale,
in massive gravity or massive bigravity, can be avoided by requiring the graviton mass to be
much higher than the present Hubble scale \cite{74}. The suggestion that the
mass of graviton could be large enough at early universe, is consistent with a small  scale factor as a solution corresponding to the initial state  of the universe before inflation \cite{3}.
For example, a  graviton with large mass $m\sim  10^{12}GeV$ at early universe has been proposed
by applying the no-boundary proposal for the quantum
cosmology of de Rham, Gabadadze and Tolley (dRGT) massive gravity theory
\cite{38},
in which
 two reasons are given for why the graviton can have large value at early universe and a negligible value today. A rather
speculative justification for the large mass of graviton  at early universe
and its small mass today was also given  based on the application
of uncertainty
principle on the universe as a single quantum system \cite{MD}. Based on
the above mentioned features, we expect
that a static Einstein universe with a sufficiently small size (well above the Planck scale) at early universe can provide us with a
sufficiently large mass of graviton such that the study of massive bigravity in the framework
of emergent universe seems reasonable and the solutions are expected to be within the regime of validity of the theory.

In spite of general relativity results where, in order to obtain static solutions we need a cosmological constant term, a positive space curvature term and a suitable perfect fluid term, the authors in \cite{26} found that in the massive gravity theory it is possible to obtain static cosmological solutions even for flat and open universes, in which only a perfect fluid term exists as a source.

The covariant massive gravity model does not show ghosts at the nonlinear level  in a certain decoupling limit which is obtained by taking $M_{Pl}\rightarrow \infty$ and $m_{graviton}\equiv m \rightarrow 0$ while keeping the scale $\Lambda_{3}=\left(m^{2}M_{Pl}\right)^{\frac{1}{3}}$ fixed, with a secondary non-dynamical reference metric \cite{17,19,20}. Besides, it turned out that  the dRGT model cannot describe  a flat universe \cite{21}.
Therefore, Hassan and Rosen tried to prove the absence of the Boulware-Deser ghosts in a Hamiltonian constraint approach \cite{22,23}, and moreover they extended the massive gravity theory beyond the dRGT model towards massive bigravity theory with two dynamical symmetric tensors $g_{\mu\nu}$ and  $f_{\mu\nu}$ having a completely symmetric role \cite{14}. Actually, considering two metrics $g_{\mu\nu}$ and  $f_{\mu\nu}$ thoroughly in a symmetric role, changes the aether-like concept of second reference metric  $f_{\mu\nu}$ in massive gravity. The cosmology of massive gravity and massive bigravity has been studied in \cite{16,24} and \cite{25}, respectively. In this work, we study the static cosmological solutions in the context of massive bigravity model \cite{14} to confirm the key role of graviton mass in obtaining the new class of static Einstein solutions subject to a perfect fluid source.
Then, we study the stability of  typical Einstein static universe in massive bigravity model. It turns
out that the graviton mass parameter plays a key role in obtaining a very small size initial scale factor avoiding the big bang singularity. We show that the obtained cosmological solutions cannot be stable in spatially flat universe with $\kappa=0$, whereas the stability is possible for open and closed universes.

We emphasize that, similar to previous works \cite{333}, in this paper merely the existence of Einstein static solution
and its stability as a background solutions against time-dependent perturbations are investigated. Obviously, this
analysis is not sufficient to establish the full behavior of all the modes present in bi-gravity because based on a purely
background analysis it is not possible to talk about Higuchi ghosts or gradient instabilities \cite{Hig,Yam} which are usually
present in bigravity. In fact, every possible solution which is called a branch can be distinguished, depending on how
the ratio of the scale factors of the metrics metrics $g_{\mu \nu}$ and $f_{\mu \nu}$ evolves. In the solutions subject to finite branches
the ratio evolves from zero towards a finite asymptotic value, whereas for the case of infinite branches the ratio
becomes infinitely large at early times and decreases with time. So far, only finite and infinite branches together with
their ghost and gradient instabilities have been extensively studied in the literature. All other branches including
bouncing cosmologies or a static universe in the asymptotic past or future are called exotic branches which have not been extensively studied in the literature \cite{Frank}. Moreover, no attempt has been done regarding the ghost and gradient
instabilities in the exotic branch. The Einstein static universe as a static universe in the
asymptotic past is also described by the exotic branch whose ghost and gradient instabilities have not been yet extracted in
the literature. Therefore, a full stability analysis of Einstein static universe in bigravity seems to be well beyond
the scope of this paper and needs a throughout investigation in another work.

The organization of this paper is as follows. In section 2, the nonlinear massive bigravity model is introduced and  the modified Friedmann equations in the presence of two isotropic and homogeneous line elements with two scale factors $a(t)$ and $b(t)$, corresponding to $g_{\mu\nu}$ and $f_{\mu\nu}$ respectively, are obtained. In section 3, we investigate the Einstein static cosmological solutions in this model to find the minimum scale factor as the initial size of the universe.
In section 4, the stability properties of the obtained Einstein static cosmological solutions are discussed in details. In section 5, the numerical behavior and the dynamics near the fixed points is studied using numerical integrations. Moreover, the phase diagrams of the system are depicted for open and closed universes in the $\left(a,\dot{a}\right)$-plane. Finally, we give a brief
conclusion, in section 6.

\section{Cosmological equations \label{Sec2}}

 Massive bigravity model is introduced by the action \cite{14}

\be
\label{Fbi1}
S_{{\rm bi}}=-\frac{M_{g}^{2}}{2}\int d^{4}x\sqrt{- {\rm det} g}R-\frac{M_{f}^{2}}{2}\int d^{4}x\sqrt{- {\rm det} f}\tilde{R}+
m^{2}M_{g}^{2}\int d^{4}x\sqrt{- {\rm det} g}\sum_{n=0}^{4}\beta_{n}e_{n}\left(\sqrt{g^{-1}f}\right)+\int d^{4}x \sqrt{- {\rm det} g}~\mathcal{L}_{m}.
\ee
Here, $g_{\mu\nu}$ and $f_{\mu\nu}$ are two dynamical metrics with corresponding Ricci scalars $R$ and $\tilde{R}$ respectively,  $\mathcal{L}_{m}~(g,\Phi)$ is the matter Lagrangian containing an scalar field $\Phi$, and the parameter $m$ describes the mass of graviton.

The square root matrix $\sqrt{g^{-1}f}$ means $\left(\sqrt{g^{-1}f}\right)^{\mu}~_{\rho}\left(\sqrt{g^{-1}f}\right)^{\rho}~_{\nu}=g^{\mu\rho}f_{\rho\nu}=X^{\mu}~_{\nu}$. The trace of this  tensor as $X^{\mu}~_{\mu}$ or $[X]$ helps us to write the following expressions for the elementary symmetric polynomials $e_{n}(X)$'s
\begin{align}\label{Fbi2}
e_{0}(X)=&1,~~e_{1}(X)=[X],~~e_{2}(X)=\frac{1}{2}\left([X]^{2}-[X^{2}]\right),\nn
e_{3}(X)=&\frac{1}{6}\left([X]^{3}-3[X][X^{2}]+2[X^{3}]\right),\nn
e_{4}(X)=&\frac{1}{24}\left([X]^{4}-6[X]^{2}[X^{2}]+3[X^{2}]^{2}+8[X][X^{3}]-6[X^{4}]\right),\nn
e_{i}(X)=&0~~{\rm for}~~i>4.
\end{align}
According to a nonlinear ADM analysis in \cite{28}, the mentioned action (\ref{Fbi1}) is ghost free and describes 7 propagating degrees of freedom. Apart from the matter coupling part, the action is invariant under the following
exchanges,

\begin{align}\label{Fbi3}
g\leftrightarrow f,~~~~~\beta_n \rightarrow \beta_4-n,~~~~~M_g \leftrightarrow M_f,~~~~m^2\rightarrow m^2M_g^2/M_f^2
\end{align}
 Now, we obtain the equations of motion   by varying the action (\ref{Fbi1})
with respect to $g_{\mu\nu}$ and $f_{\mu\nu}$, respectively as
\begin{align}
\label{Fbi4}
0=R_{\mu\nu}-\frac{1}{2}g_{\mu\nu}R+\frac{m^{2}}{2}\sum_{n=0}^{3}(-1)^n \beta_{n} \left[g_{\mu\lambda}Y_{(n)\nu}^{\lambda}\left(\sqrt{g^{-1}f}\right)+g_{\nu\lambda}Y_{(n)\mu}^{\lambda}\left(\sqrt{g^{-1}f}\right)\right]
-\frac{T_{\mu\nu}}{M_{g}^2},
\end{align}
and
\begin{align}\label{Fbi5}
0=\tilde{R}_{\mu\nu}-\frac{1}{2}f_{\mu\nu}\tilde{R}+\frac{m^{2}}{2M_{*}^2}\sum_{n=0}^{3}(-1)^n \beta_{4-n} \left[f_{\mu\lambda}Y_{(n)\nu}^{\lambda}\left(\sqrt{f^{-1}g}\right)+f_{\nu\lambda}Y_{(n)\mu}^{\lambda}\left(\sqrt{f^{-1}g}\right)\right],
\end{align}
where we have introduced the ratio
\begin{align}\label{Fbi6}
M_{*}^{2}\equiv \frac{M_f^2}{M_g^2}.
\end{align}
Meanwhile, the matrices $Y_{(n)\mu}^{\lambda}\left(X\right)$ are given by
\begin{align}\label{Fbi7}
Y_{(0)}\left(X\right)=&1,~~~Y_{(1)}\left(X\right)=X-1\left[X\right],\nn Y_{(2)}\left(X\right)=&X^2-X\left[X\right]+\frac{1}{2}1\left(\left[X\right]^2-\left[X^2\right]\right),\nn
Y_{(2)}\left(X\right)=&X^3-X^2\left[X\right]+\frac{1}{2}X\left(\left[X\right]^2-\left[X^2\right]\right)-\frac{1}{6}1\left(\left[X\right]^3-
3\left[X\right]\left[X^2\right]+2\left[X^3\right]\right).
\end{align}
As a consequence of the covariant conservation of $T_{\mu\nu}$ and also the Bianchi identity, the equation (\ref{Fbi4}) leads to the Bianchi constraint for the metric $g_{\mu\nu}$
\be\label{Fbi8}
0=\nabla^{\mu} \sum_{n=0}^{3}\left(-1\right)^{n}\beta_{n} \left[g_{\mu\lambda}Y_{(n)\nu}^{\lambda}\left(\sqrt{g^{-1}f}\right)+g_{\nu\lambda}Y_{(n)\mu}^{\lambda}\left(\sqrt{g^{-1}f}\right)\right].
\ee
Similarly, the equation (\ref{Fbi5}) gives us the Bianchi constraint corresponding to the metric $f_{\mu\nu}$
\be\label{Fbi9}
0=\tilde{\nabla}^{\mu} \sum_{n=0}^{3}\left(-1\right)^{n}\beta_{4-n} \left[f_{\mu\lambda}Y_{(n)\nu}^{\lambda}\left(\sqrt{f^{-1}g}\right)+f_{\nu\lambda}Y_{(n)\mu}^{\lambda}\left(\sqrt{f^{-1}g}\right)\right],
\ee
where $\tilde{\nabla}^{\mu}$ indicates the covariant derivatives with respect to the metric $f_{\mu\nu}$. Two above Bianchi constraints are equivalent which is a direct result of the invariance of the interaction term under the general coordinate transformations of the two metrics, so we just consider the constraint (\ref{Fbi8}).
We consider a Friedmann-Robertson-Walker (FRW) universe with three-dimensional spatial curvature $\kappa=0,\pm1$  for both metrics which exhibit spatial isotropy and homogeneity
\be
\label{Fbi10}
ds_{g}^{2}=-dt^{2}+a(t)^{2}\left(\frac{dr^{2}}{1-\kappa r^{2}}+r^{2}d\theta^{2}+r^{2}\sin^{2}\theta d\varphi^{2}\right),
\ee
\be
\label{Fbi11}
ds_{f}^{2}=-c(t)^{2}dt^{2}+b(t)^{2}\left(\frac{dr^{2}}{1-\kappa r^{2}}+r^{2}d\theta^{2}+r^{2}\sin^{2}\theta d\varphi^{2}\right).
\ee
For the metrics (\ref{Fbi10}) and (\ref{Fbi11}), the Bianchi constraints (\ref{Fbi8}) or (\ref{Fbi9}) gives (The mass interaction term is invariant under the diagonal subgroup of the general coordinate transformations of the two metrics and hence these Bianchi constraints are equivalent)
\be\label{Fbi12}
c(t)=\frac{\dot{b}}{\dot{a}}~,
\ee
where $c(t)$ is the lapse function of $f_{\mu\nu}$ metric . This is an important result for the next calculations. The modified Friedmann equation and the modified acceleration equation corresponding to the metric $g_{\mu\nu}$ are as follows
\be\label{Fbi13}
-3\left(\frac{\dot{a}}{a}\right)^{2}-\frac{3\kappa}{a^2}+m^{2}\left[\beta_{0}+3\beta_{1}\frac{b}{a}+3\beta_{2}\frac{b^2}{a^2}+
\beta_{3}\frac{b^3}{a^3}\right]=-\frac{\rho}{M_{g}^2},
\ee
\be\label{Fbi14}
-2\frac{\ddot{a}}{a}-\frac{\dot{a}^2}{a^2}-\frac{\kappa}{a^2}+m^{2}\left[\beta_{0}+2\beta_{1}\left(\frac{b}{a}+\frac{\dot{b}}{\dot{a}}\right)+
\beta_{2}\left(\frac{b^2}{a^2}+\frac{2b\dot{b}}{a\dot{a}}\right)+\beta_{3}\frac{b^2\dot{b}}{a^2\dot{a}}\right]=\frac{P}{M_{g}^2}.
\ee
Note that although the ordinary Friedmann and accelerating equations are recovered in the limit $m^{2}\rightarrow 0$, however the cosmological solutions will not be well defined in this limit.
Consequently, the $f_{\mu\nu}$ equation of motion (\ref{Fbi5}) is obtained as
\be\label{Fbi15}
-3\left(\frac{\dot{a}}{b}\right)^{2}-\frac{3\kappa}{b^2}+\frac{m^2}{M_{*}^2}\left[\beta_{4}+3\beta_{3}\frac{a}{b}+3\beta_{2}\frac{a^2}{b^2}+
\beta_{1}\frac{a^3}{b^3}\right]=0,
\ee
and
 \be\label{Fbi16}
-2\frac{\dot{a}\ddot{a}}{b\dot{b}}-\frac{\dot{a}^2}{b^2}-\frac{\kappa}{b^2}+\frac{m^2}{M_{*}^2}\left[\beta_{4}+\beta_{3}
\left(2\frac{a}{b}+\frac{\dot{a}}{\dot{b}}\right)+\beta_{2}\left(\frac{a^2}{b^2}+\frac{2a\dot{a}}{b\dot{b}}\right)+
\beta_{1}\frac{a^{2}\dot{a}}{b^{2}\dot{b}}\right]=0.
\ee
Let us now explain about the matter source. Assuming an equation of state of the normal form $P(t)=\omega \rho(t)$ in the minimal coupling of the matter to gravity, the continuity equation is
\be\label{Fbi17}
\dot{\rho}+3H\left(1+\omega\right)\rho=0,
\ee
where $H=\frac{\dot{a}}{a}$ is the Hubble parameter of the scale factor $a$.

\section{The Einstein static solution in massive bigravity \label{Sec3}}

In this section we will study the Einstein static solution in massive bigravity. The following conditions describe the Einstein static solutions of massive bigravity modified Friedmann equations
\be\label{Fbi18}
\dot{a}=\ddot{a}=H=0,~~~~~~\dot{b}=\ddot{b}=K=0,
\ee
where $K$ is the Hubble parameter of the scale factor $b$. Having considered (\ref{Fbi18}), we can reduce the continuity equation (\ref{Fbi17}) as
\be\label{Fbi19}
\dot{\rho}=0.
\ee
As a result, we can assume
\be\label{Fbi20}
a=a_{Es},~~~~H\left(a_{Es}\right)=0, ~~~   b=b_{Es},~~{\rm{}}~~K\left(b_{Es}\right)=0,
\ee
and define
\be\label{Fbi21}
\frac{b_{Es}}{a_{Es}}=\gamma_{Es}.
\ee
Leaving the quantity $c(t)$ in the Bianchi constraint (\ref{Fbi12}) undetermined, we can not continue our calculation explicitly. Fortunately, the relations (\ref{Fbi15}) and (\ref{Fbi16}) help us to handle this problem. Imposing the conditions (\ref{Fbi18}) on equations (\ref{Fbi15}) and (\ref{Fbi16}), we can write

\be\label{Fbi22}
-\frac{3\kappa}{\bar{b}^2}+\frac{m^2}{M_{*}^{2}}\left(\beta_{4}+3\beta_{3}\bar{\gamma}^{-1}+3\beta_{2}\bar{\gamma}^{-2}+
\beta_{1}\bar{\gamma}^{-3}\right)=0,
\ee
\be\label{Fbi23}
\frac{3\kappa}{\bar{b}^2}-3\frac{m^2}{M_{*}^{2}}\left(\beta_{4}+\beta_{3}\left(2\bar{\gamma}^{-1}+\bar{c}^{-1}\right)+\beta_{2}\left(\bar{\gamma}^{-2}
+2\bar{\gamma}^{-1}\bar{c}^{-1}\right)
+\frac{\beta_{1}}{\bar{\gamma}^{2}\bar{c}}\right)=0,
\ee
where we have assumed $\bar{b}$, $\bar{c}$ and $\bar{\gamma}$ as  $b_{Es}$, $c_{Es}$ and $\gamma_{Es}$, respectively. Subtracting two above equations, gives
\be\label{Fbi24}
\bar{c}=\frac{-3\bar{\gamma}\left(\beta_{1}+\bar{\gamma}\left(2\beta_{2}+\bar{\gamma}\beta_{3}\right)\right)}{-\beta_{1}+\bar{\gamma}^{2}\left(3\beta_{3}
+2\bar{\gamma}\beta_{4}\right)}.
\ee
In the present model, the interaction between
two metrics $g_{\mu \nu}$ and $f_{\mu \nu}$ is  described just by the trace of $\left(\sqrt{g^{-1}f}\right)^{\mu}~_{\rho}$, hence this interaction is called  as the {\it minimal} interaction and the corresponding model is called  as  the {\it
minimal} model. This minimal model, as a simplest but non-trivial case, was proposed in \cite{29}. In the non-minimal models, the
calculations become more complicated than the minimal models. Hence, just for simplicity, we only investigate the minimal model which is described
by the following interaction term \cite{29}, \cite{Bamba}

\be\label{Fbi25}
m^{2}M_{g}^{2}\int d^{4}x\sqrt{- {\rm det} g}\sum_{n=0}^{4}\beta_{n}e_{n}\left(\sqrt{g^{-1}f}\right)=m^{2}M_{g}^{2}\int d^{4}x\sqrt{- {\rm det} g}
\left(3-{\rm tr}\sqrt{g^{-1}f}+{\rm det}\sqrt{g^{-1}f}\right),
\ee
in which we have put $\beta_{0}=3$, $\beta_{1}=-1$ and $\beta_{4}=1$. As a result, the relation (\ref{Fbi24}) reads
\be\label{Fbi26}
\bar{c}_{\rm{min}}=\frac{3 \bar{\gamma}}{1+2 \bar{\gamma}^{3}}.
\ee
Having considered the conditions (\ref{Fbi18}), we rewrite (\ref{Fbi13}) and (\ref{Fbi14}) as follows
\be\label{Fbi27}
-\frac{\kappa}{\bar{a}^{2}}=\frac{-\bar{\rho}}{3M_{g}^{2}}-\frac{m^{2}}{3}\left(\beta_{0}+3\beta_{1}\bar{\gamma}+3\beta_{2}\bar{\gamma}^{2}+
\beta_{3}\bar{\gamma}^{3}\right),
\ee
\be\label{Fbi28}
-\frac{\kappa}{\bar{a}^{2}}=-m^{2}\left(\beta_{0}+2\beta_{1}\left(\bar{\gamma}+\bar{c}\right)+\beta_{2}\left(\bar{\gamma}^{2}
+2\bar{\gamma}\bar{c}\right)+\beta_{3}\bar{\gamma}^{2}\bar{c}\right)+\frac{\omega \bar{\rho}}{M_{g}^{2}}.
\ee
Combination of these equations gives
\be\label{Fbi29}
\bar{\rho}=\frac{m^{2}M_{g}^{2}}{\left(\omega+\frac{1}{3}\right)}\left(\frac{2\beta_{0}}{3}+\beta_{1}\left(\bar{\gamma}+2\bar{c}\right)+
2\beta_{2}\bar{\gamma}\bar{c}+\beta_{3}\left(\bar{\gamma}^{2}\bar{c}-\frac{\bar{\gamma}^{3}}{3}\right)\right).
\ee
Inserting (\ref{Fbi24}) into (\ref{Fbi29}), we can find
\begin{align}\label{Fbi30}
\bar{\rho}=&
\frac{m^{2}M_{g}^{2}}{\left(\omega+\frac{1}{3}\right)}\times
.\nn&
\frac{\beta_{0}\left(-2\beta_{1}+6\bar{\gamma}^{2}\beta_{3}+4\bar{\gamma}^{3}\beta_{4}\right)-\bar{\gamma}\left(21\beta_{1}^{2}+
\bar{\gamma}\beta_{1}\left(54\beta_{2}+\bar{\gamma}\left(17\beta_{3}-6\bar{\gamma}\beta_{4}\right)\right)+
2\bar{\gamma}^{2}\left(18\beta_{2}^{2}+18\bar{\gamma}\beta_{2}\beta_{3}+\bar{\gamma}^{2}\beta_{3}\left(6\beta_{3}+
\bar{\gamma}\beta_{4}\right)\right)\right)}{-3\beta_{1}+9\bar{\gamma}^{2}\beta_{3}+6\bar{\gamma}^{3}\beta_{4}}.
\end{align}
The minimal case of this theory takes the following simplification
\be\label{Fbi31}
\bar{\rho}_{\rm{min}}=\frac{m^{2}M_{g}^{2}}{\left(\omega+\frac{1}{3}\right)}\left(2+\bar{\gamma}\left(-1-\frac{6}{1+
2\bar{\gamma}^{3}}\right)\right).
\ee
Referring to (\ref{Fbi27}),  we can extract the squared Einstein static scale factor for the ordinary case as
\bea\label{Fbi32}
\bar{a}^{2}&=&\kappa (3 \omega +1) \left(\beta_1-\bar{\gamma} ^2 \left(3 \beta_3+2 \bar{\gamma } \beta_4\right)\right)
\times\\\nonumber&&
[m^2 ((\omega +1) \beta_0 t(\beta_1-\bar{\gamma} ^2 (3 \beta_3+2 \bar{\gamma}
\beta_4))+\bar{\gamma}  (-(\beta_3 (3 (\omega -1) \beta_3+2 \bar{\gamma}  \omega \beta_4) \bar{\gamma} ^2+\\\nonumber&& \beta_2 (9 (\omega -1) \beta_3+2 \bar{\gamma}  (3 \omega +1)
\beta_4) \bar{\gamma} -12 \beta_2^2) \bar{\gamma} ^2+\beta_1 ((3 \omega +19) \beta_2-\bar{\gamma}  ((8 \omega -3)\beta_3+2 \bar{\gamma}  (3 \omega +2) \beta_4)) \bar{\gamma} +(3
\omega +8)\beta_1^2))]^{-1},
\eea
and the squared Einstein static scale factor for the minimal case
as
\be\label{Fbi33}
\bar{a}_{\rm{min}}^{2}=-\frac{3\kappa}{m^{2}\left(3-3\bar{\gamma}+\frac{2+\bar{\gamma}\left(-1-\frac{6}{1+
2\bar{\gamma}^{3}}\right)}{\omega+\frac{1}{3}}\right)}.
\ee
It is obvious that the size of the critical scale factor corresponding to the minimum scale factor of Einstein static universe is affected by the mass of gravitons, and that a non-vanishing small size of non-singular universe requires $\kappa\neq 0$ and large mass gravitons.
\\
\\
\textbf{Case}~\textbf{1}~\textbf{:}~\textbf{ $\kappa=1$}\\
\\
The requirement $\bar{a}^2_{\rm{min}}>0$   gives the inequality
\be\label{Fbi34}
\frac{3}{3-3\bar{\gamma}+\frac{2+\bar{\gamma}\left(-1-\frac{6}{1+
2\bar{\gamma}^{3}}\right)}{\omega+\frac{1}{3}}}<0,
\ee
 according to which the allowed ranges of  $\bar{\gamma}>0$ and $\omega$
are obtained in Table 1.
\vspace{5mm}
\begin{center}
{\scriptsize{ Table 1: }}\hspace{-2mm} {\scriptsize Allowed ranges of $\omega$ and $\bar{\gamma}$ for the case $\kappa=1$, by the requirement $\bar{a}^2_{\rm{min}}>0$.}\\
    \begin{tabular}{|l| l |l |  p{800mm} }
    \hline
   {\footnotesize$~~~~~~\omega$ }& ~~{\footnotesize~~~~~ $\bar{\gamma}$ }  \\ \hline
{\footnotesize ~$\left(-\infty,-0.7\right)$} & ~~{\footnotesize $\left[1.6,+\infty\right)$} \\\hline
 {\footnotesize ~$\left(-\frac{1}{3},+\infty\right)$} & ~{\footnotesize  ~$\left[1.6,+\infty\right)$}
\\ \hline
    \end{tabular}
\end{center}
\textbf{Case}~\textbf{2}~\textbf{:}~\textbf{ $\kappa=-1$}\\ \\
For the open universe  we obtain
\be\label{Fbi35}
\frac{-3}{3-3\bar{\gamma}+\frac{2+\bar{\gamma}\left(-1-\frac{6}{1+
2\bar{\gamma}^{3}}\right)}{\omega+\frac{1}{3}}}<0.
\ee
and the allowed ranges of $\omega$ and $\bar{\gamma}$ are given in Table
2.
\vspace{5mm}
\begin{center}
{\scriptsize{ Table 2: }}\hspace{-2mm} {\scriptsize Allowed ranges of $\omega$ and $\bar{\gamma}$ for the case $\kappa=-1$, by the requirement $\bar{a}^2_{\rm{min}}>0$.}\\
    \begin{tabular}{|l| l |l |  p{800mm} }
    \hline
   {\footnotesize$~~~~~~\omega$ }& ~~{\footnotesize~~ $\bar{\gamma}$ }  \\ \hline
{\footnotesize ~$\left(-\infty,-\frac{1}{3}\right)$} & ~~~{\footnotesize $\left[0.3,1\right)$} \\\hline
{\footnotesize ~$\left(-\frac{1}{3},+\infty\right)$} & ~~~{\footnotesize $\left(0,0.3\right]$} \\\hline
    \end{tabular}
\end{center}
On the other hand,  imposing the nonvanishing sector of weak energy condition
as $\bar{\rho}_{\rm{min}}>0$
leads to the following results
\begin{equation}
\omega>-\frac{1}{3}~~~~\Longrightarrow ~~~~ 0<\bar{\gamma}\leq0.3
\end{equation}
\begin{equation}
\omega<-\frac{1}{3}~~~~\Longrightarrow ~~~~ \bar{\gamma}>0.3
\end{equation}
Considering these results, we obtain the following ranges of $\omega$ and $\bar{\gamma}$   given in Table 3
\begin{center}
\vspace{5mm}
{\scriptsize{ Table 3: }}\hspace{-2mm} {\scriptsize Allowed ranges for $\omega$ and $\bar{\gamma}$ for the case $\kappa=1$, by the requirements $\bar{a}^2_{\rm{min}}>0$ and  $\bar{\rho}_{\rm{min}}>0$.}\\
    \begin{tabular}{|l| l |l |  p{800mm} }
    \hline
   {\footnotesize$~~~~~~~~\omega$ }& ~~{\footnotesize~~~ $\bar{\gamma}$ }  \\ \hline
{\footnotesize ~$\left(-\infty,-0.7\right)$} & ~{\footnotesize $\left[1.6,+\infty\right)$} \\ \hline
    \end{tabular}
\end{center}
for the case $\kappa=1$,
and the following ranges of $\omega$ and $\bar{\gamma}$ given in Table 4
\begin{center}
\vspace{5mm}
{\scriptsize{ Table 4: }}\hspace{-2mm} {\scriptsize Allowed ranges of $\omega$ and $\bar{\gamma}$ for the case $\kappa=-1$, by the requirements $\bar{a}^2_{\rm{min}}>0$ and  $\bar{\rho}_{\rm{min}}>0$.}\\
    \begin{tabular}{|l| l |l |  p{800mm} }
    \hline
   {\footnotesize$~~~~~~\omega$ }& ~~{\footnotesize~~ $\bar{\gamma}$ }  \\ \hline
{\footnotesize ~$\left(-\infty,-\frac{1}{3}\right)$} & ~{\footnotesize $\left(0.3,1\right)$} \\\hline
{\footnotesize ~$\left(-\frac{1}{3},+\infty\right)$} & ~{\footnotesize $\left(0,0.3\right]$} \\\hline
    \end{tabular}
\end{center}
for the case $\kappa=1$.

\section{Stability of critical points \label{Sec4}}
In this section, we discuss the stability of the critical points (\ref{Fbi18}) named $\bar{a}$, $\bar{H}$, $\bar{b}$ and $\bar{K}$.
Let us use the Bianchi constraint (\ref{Fbi12}) to write
\be\label{Fbi36}
c=\frac{bK}{aH}.
\ee
Taking time derivative, one obtain
\be\label{Fbi37}
\dot{c}=\frac{\dot{b}K}{aH}+\frac{b\dot{K}}{aH}-\frac{bK\dot{a}}{a^{2}H}-\frac{bK\dot{H}}{aH^{2}}.
\ee
In the case of minimal massive bigravity models, the first and second Friedmann equations (\ref{Fbi13}), (\ref{Fbi14}), (\ref{Fbi15}) and (\ref{Fbi16}) reduce to
\be\label{Fbi38}
-3\left(\frac{\dot{a}}{a}\right)^{2}-\frac{3\kappa}{a^{2}}+3m^{2}\left(1-\frac{b}{a}\right)=-\frac{\rho}{M_{g}^{2}},
\ee
and
\be\label{Fbi39}
-2\frac{\ddot{a}}{a}-\left(\frac{\dot{a}}{a}\right)^{2}-\frac{\kappa}{a^{2}}+m^{2}\left(3-2\left(\frac{b}{a}+\frac{\dot{b}}{\dot{a}}\right)\right)
=\frac{\rho \omega}{M_{g}^{2}},
\ee
which correspond to the metric $g_{\mu\nu}$. And also for the metric $f_{\mu\nu}$ we have
\be\label{Fbi40}
-3\left(\frac{\dot{a}}{b}\right)^{2}-\frac{3\kappa}{b^{2}}+\frac{m^{2}}{M_{*}^{2}}\left(1-\frac{a^3}{b^3}\right)=0,
\ee
\be\label{Fbi41}
-2\frac{\ddot{a}\dot{a}}{b\dot{b}}-\left(\frac{\dot{a}}{b}\right)^{2}-\frac{\kappa}{b^{2}}+\frac{m^{2}}{M_{*}^2}
\left(1-\frac{a^{2}\dot{a}}{b^{2}\dot{b}}\right)=0.
\ee
It should be mentioned that for simplicity we have not used index (min) for $a$, $b$, $c$, $\rho$ and $\gamma$, representing these quantities in minimal massive bigravity cosmological model.
To further analyze, we have to extract time derivatives of the quantities $H$, $K$, $L$ ($L=\frac{\dot{c}}{c}$), $a$, $b$ and $c$ to construct the following eigenvalue relation
\begin{align}
\label{Fbibi42}
 \frac{d}{dt}\left(
 \begin{array}{c}
 H \\
  K \\
  L \\
  a\\
 b\\
  c
  \end{array}
  \right)=M
  \left(
  \begin{array}{c}
  H  \\
  K \\
  L \\
  a\\
b\\
  c
  \end{array}\right),
  \end{align}
where $M=\frac{\partial y}{\partial x}$ (a $6\times6$ ~Jacobian matrix) in which $y=\left(\dot{H},\dot{K},\dot{L},\dot{a},\dot{b},\dot{c}\right)$ and $x=\left(H,K,L,a,b,c \right)$.
By means of equations (\ref{Fbi36})-(\ref{Fbi41}) and a troublesome calculation we obtain
\be\label{Fbi43}
\dot{H}=-\frac{3}{2}H^{2}\left(1+\omega\right)-\frac{\alpha}{2a^{2}}\left(1+3\omega\right)+m^{2}\left(\frac{3}{2}-\frac{b}{a}-c-\frac{3\omega b}{2a}+\frac{3\omega}{2}\right),
\ee
\be\label{Fbi44}
\dot{K}=-\frac{3}{2}K^{2}+K L-\frac{\alpha c^{2}}{2b^{2}}+\frac{m^{2}}{2M_{*}^{2}}\left(c^{2}-\frac{a^{2}c}{b^{2}}\right),
\ee
and
\begin{align}\label{Fbi45}
\dot{L}=&H^{2}-2HK+K^{2}-L^{2}
-
\nn&
\frac{\left(H-2K\right)\left(3a^{2}H\left(H^{2}-m^{2}\right)\left(1+\omega\right)
+H\alpha \left(1+3\omega\right)+m^{2}ab\left(2\left(H+K\right)+3H\omega\right)\right)}{2a^{2}H^{2}}+\nn& \frac{\left(3a^{2}H\left(H^{2}-m^{2}\right)\left(1+\omega\right)+H\alpha \left(1+3\omega\right)+m^{2}ab \left(2\left(H+K\right)+3H\omega\right)\right)^{2}}{2a^{4}H^{4}}+\nn&
\frac{\left(2H-K\right)\left(m^{2}\left(a^{3}H-b^{3}K\right)+b\left(a^{2}H^{2}\left(3K-2L\right)+K \alpha\right)M_{*}^{2}\right)}{2a^{2}bH^{2}M_{*}^{2}}-\nn&
\frac{\left(3a^{2}H\left(H^{2}-m^{2}\right)\left(1+\omega\right)+H\alpha \left(1+3\omega\right)+m^{2}ab \left(2\left(H+K\right)
+3H\omega\right)\right)}{2a^{4}bH^{4}M_{*}^{2}}\times
\nn&
\left(m^{2}\left(a^{3}H-b^{3}K\right)+b\left(a^{2}H^{2}\left(3K-2L\right)+K\alpha\right)
M_{*}^{2}\right).
\end{align}
Additionally, we have
\be\label{Fbi46}
\dot{a}=aH,
\ee
\be\label{Fbi47}
\dot{b}=bK,
\ee
\be\label{Fbi48}
\dot{c}=cL.
\ee
As mentioned in (\ref{Fbi18}), (\ref{Fbi20}) and (\ref{Fbi21}), the Einstein static solution corresponds to the fixed points, $\bar{H}=\bar{K}=\bar{L}=0$ , $\bar{a}$ and $\bar{b}$. Considering these fixed points in the relation (\ref{Fbi36}), we can find the following behavior for the fixed point $\bar{c}$
\be\label{Fbi49}
\bar{c}\rightarrow\frac{\bar{b}}{\bar{a}}=\bar{\gamma}.
\ee
The eigenvalue equation with the eigenvalues $\lambda$ corresponding to (\ref{Fbibi42}) has the following form
\be\label{Fbi50}
0=\lambda^{6}+g_{5}\lambda^{5}+g_{4}\lambda^{4}+g_{3}\lambda^{3}+g_{2}\lambda^{2}+g_{1}\lambda+g_{0}.
\ee
The stability analysis of the presented solutions can be performed by requiring all the  eigenvalues $\lambda$ to be negative \footnote{The stability of fixed points of a system of constant coefficient linear differential equations of first order can be analyzed using the eigenvalues of the corresponding matrix. An autonomous system $x'=Ax$, where $x(t)\in R_n$ and $A$ is an $n\times
n$ matrix with real entries, has a constant solution $x(t)=0$.
This solution is asymptotically stable as $t \rightarrow \infty$ ("in the future") if and only if for all eigenvalues $\lambda$ of $A$, $Re(\lambda) < 0$. Similarly, it is asymptotically stable as $t \rightarrow -\infty$ ("in the past") if and only if for all eigenvalues $\lambda$ of $A$, $Re(\lambda) > 0$.}. This is because  all the eigenmodes with negative eigenvalues gradually disappear and thus the perturbation is damped. In order to investigate the
condition for which all the  eigenvalues $\lambda$ are negative, we benefit of the
following procedure.  The equation (\ref{Fbi50}) is a special case of the general form of the eigenvalue function expansion of an arbitrary matrix ($\rm{n\times n}$) as follows
\be\label{Fbi51}
(-\lambda)^{n}+\rm{trM}(-\lambda)^{n-1}+...+\rm{detM}=F_{M}(\lambda).
\ee
Comparing (\ref{Fbi50}) and (\ref{Fbi51}) for $n=6$, and considering the
fact that $g_i>0~ (i=1,...,5)$ is the requirement
of having  all the eigenvalues to be negative, we
obtain the following condition \be\label{Fbi52}
g_{_{5}}=-\rm{trM}>0 \Longrightarrow \rm{trM}<0.
\ee
By imposing the Einstein static condition, we can obtain
\be\label{Fbi53}
\rm{trM}=\rm{\lim}_{\bar{H}\rightarrow 0}\frac{1}{\bar{H}}\left(5m^{2}\bar{\gamma}^{2}+\frac{m^{2}}{2M_{*}^{2}}\left(2\bar{\gamma}^{2}-\frac{1}{\bar{\gamma}}\right)-3m^{2}\left(1+\omega\right)
+\frac{3\omega\kappa}{\bar{a}^{2}}+3m^{2}\omega \bar{\gamma}\right).
\ee
 The above expression goes to infinity in the limit of ${\bar{H}}\rightarrow 0$, however we are interested in determining the signature of this infinity as positive infinity or negative one. In doing so, we should plot the three dimensional diagram of the parenthesis in equation (\ref{Fbi53}) in which there are two variables $\omega$ and $\bar{\gamma}$. Since equation (\ref{Fbi53}) has the same form in both cases $\kappa=1$ and $\kappa=-1$, we can write the following result
\vspace{5mm}
\begin{center}
{\scriptsize{ Table 5: }}\hspace{-2mm} {\scriptsize Allowed ranges of $\omega$ and $\bar{\gamma}$ for the cases $\kappa=1$ and $\kappa=-1$.}\\
    \begin{tabular}{|l| l |l |  p{800mm} }
    \hline
   {\footnotesize$~~~~~~~\omega$ }& ~~{\footnotesize~~ $\bar{\gamma}$ }  \\ \hline
{\footnotesize $[-0.27,+\infty)$} & ~~{\footnotesize $(0,0.3]$} \\ \hline
 {\footnotesize ~~$(-12,-3)$} & ~~{\footnotesize $(1.6,10)$ } \\ \hline
\end{tabular}
\end{center}
This is in accordance with the allowed ranges in Table 3 and Table 4. As a result, the $\omega$ intervals  $[-0.27,+\infty)$ and  $(-12,-3)$ correspond to $\bar{b}<\bar{a}$ and $\bar{b}>\bar{a}$ respectively, in order to have some stable cosmological solutions. This shows an interesting result that a competition between two scale factors of massive bigravity model corresponds to a competition between different $\omega$ parameter
spaces.

\section{Numerical behavior of the Cosmological dynamics in minimal massive bigravity theory \label{Sec4}}
Now, we have found the allowed ranges of variables $\omega$ and $\bar{\gamma}$ in minimal massive bigravity model which result in the stable Einstein static state.
Here, we make a brief quantitative study of the cosmological dynamics of Einstein static universe for the closed and open universe.\\

Considering the minimal massive bigravity, the modified Friedmann equations (\ref{Fbi13}), (\ref{Fbi15}) and the modified acceleration equations (\ref{Fbi14}),
(\ref{Fbi16}) with the coefficients $\beta_{1}=-1$, $\beta_{0}=3$ and $\beta_{4}=1$  for $g_{\mu\nu}$,  $f_{\mu\nu}$ metrics become respectively as
\be\label{Fbi54}
-3\left(\frac{\dot{a}}{a}\right)^{2}-\frac{3\kappa}{a^{2}}+3m^{2}\left(1-\frac{b}{a}\right)=-\frac{\rho}{M_{g}^{2}},
\ee
\be\label{Fbi55}
-2\frac{\ddot{a}}{a}-\frac{\dot{a}^{2}}{a^{2}}-\frac{\kappa}{a^{2}}+m^{2}\left(3-2\left(\frac{b}{a}+\frac{\dot{b}}{\dot{a}}\right)\right)=
\frac{P}{M_{g}^{2}},
\ee
 and
\be\label{Fbi56}
-3\left(\frac{\dot{a}}{b}\right)^{2}-\frac{3\kappa}{b^{2}}+\frac{m^{2}}{M_{*}^{2}}\left(1-\frac{a^{3}}{b^{3}}\right)=0,
\ee
\be\label{Fbi57}
-\frac{2\dot{a}\ddot{a}}{b\dot{b}}-\frac{\dot{a}^{2}}{b^{2}}-\frac{\kappa}{b^{2}}+\frac{m^{2}}{M_{*}^{2}}
\left(1-\frac{a^{2}\dot{a}}{b^{2}\dot{b}}\right)=0.
\ee

To further analyze the reduced Friedmann equation we need the variable $\rho$ in terms of the ratio of two scale factors  $\frac{b}{a}=\gamma$.
Combination of the equations (\ref{Fbi54}) and (\ref{Fbi56}) gives
\be\label{Fbi58}
\rho=m^{2}\left(\frac{M_{g}^{2}}{M_{*}^{2}}\left(\gamma^{2}-\gamma^{-1}\right)-3+3\gamma\right).
\ee
\textbf{Case}~\textbf{1}~\textbf{:}~\textbf{ $\kappa=1$}\\

{ Considering (\ref{Fbi54}), we can extract the energy density
{\be\label{Fbi59}
\rho=M_{g}^{2}\left(3H^{2}+\frac{3}{a^{2}}+3m^{2}\left(\gamma-1\right)\right).
\ee}
{Using (\ref{Fbi58}) and (\ref{Fbi59}), we have the following relation
\be\label{Fbi60}
3H^{2}+\frac{3}{a^{2}}+m^{2}\left(3\gamma-3-\frac{1}{M_{*}^{2}}\left(\gamma^{2}-\frac{1}{\gamma}\right)+\frac{1}{M_{g}^{2}}
\left(3-3\gamma\right)\right)=0.
\ee
Without loss of generality, here we assume $M_{g}=M_{f}=M_{Pl}$ for more
simplicity. As a result the above relation reads
as\be\label{Fbi61}
3H^{2}+\frac{3}{a^{2}}+m^{2}\left(-\gamma^{2}+3\gamma-3+\frac{1}{\gamma}+\frac{1}{M_{Pl}^{2}}\left(3-3\gamma\right)\right)=0.
\ee
This determines $\gamma$ as a very complicate function of $a$,  $H$,  $M_{Pl}$ and the graviton mass $m$.} 
}
Moreover, $\frac{\dot{b}}{\dot{a}}$ becomes
\be\label{Fbi62}
\frac{\dot{b}}{\dot{a}}=\frac{\dot{\gamma}}{H}+\gamma.
\ee
Eq. (\ref{Fbi61}) can help us to extract $\dot{\gamma}$, but it is too complicated to be mentioned here. Combining (\ref{Fbi55}) and { (\ref{Fbi59}) with (\ref{Fbi62}), we obtain the second Friedmann equation}
\be\label{Fbi63}
2\frac{\ddot{a}}{a}+\frac{\dot{a}^{2}}{a^{2}}\left(1+3\omega\right)+\frac{1}{a^{2}}\left(1+3\omega\right)+m^{2}\left(-3+4\gamma+
\frac{2\dot{\gamma}}{\frac{\dot{a}}{a}}+3\omega\left(\gamma-1\right)\right)=0.
\ee
According to the stability results of fixed points in  Tables 3 and 5 and the allowed ranges of  $\omega$ and $\bar{\gamma}$, the universe can stay at the stable state (\ref{Fbi33}) for the scale factor $a$ past eternally and also may undergo some indefinite, non-singular oscillations, as shown in Fig.1, which typically shows the avoidance of big bang singularity properly
for closed universe at phantom dominant era. Due to the correspondence between the scale factors $a$ and $b$ through $\gamma$,  the stability
of scale factor $a$ indicates for the stability of   scale
factor $b$. \\ \\
\begin{figure*}[ht]
  \centering
  \includegraphics[width=2.5in]{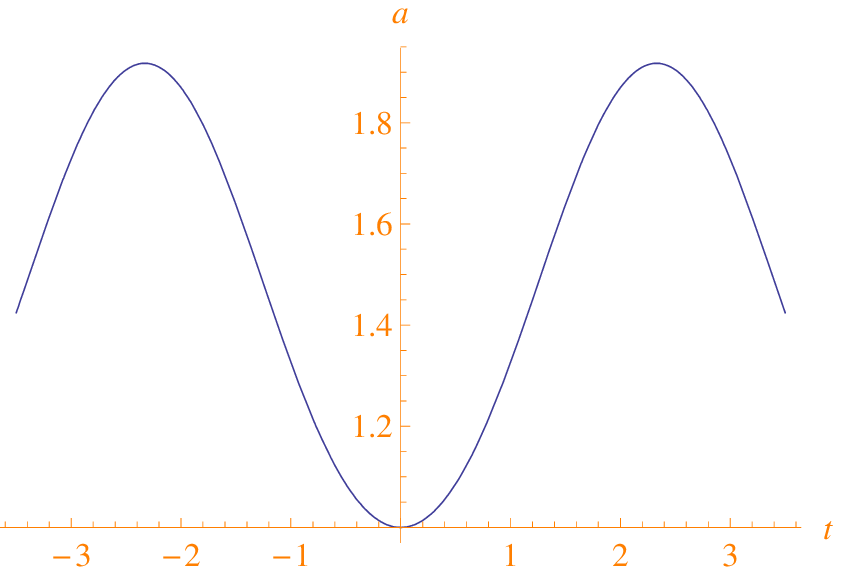}\hspace{2cm}
  \includegraphics[width=1.6in]{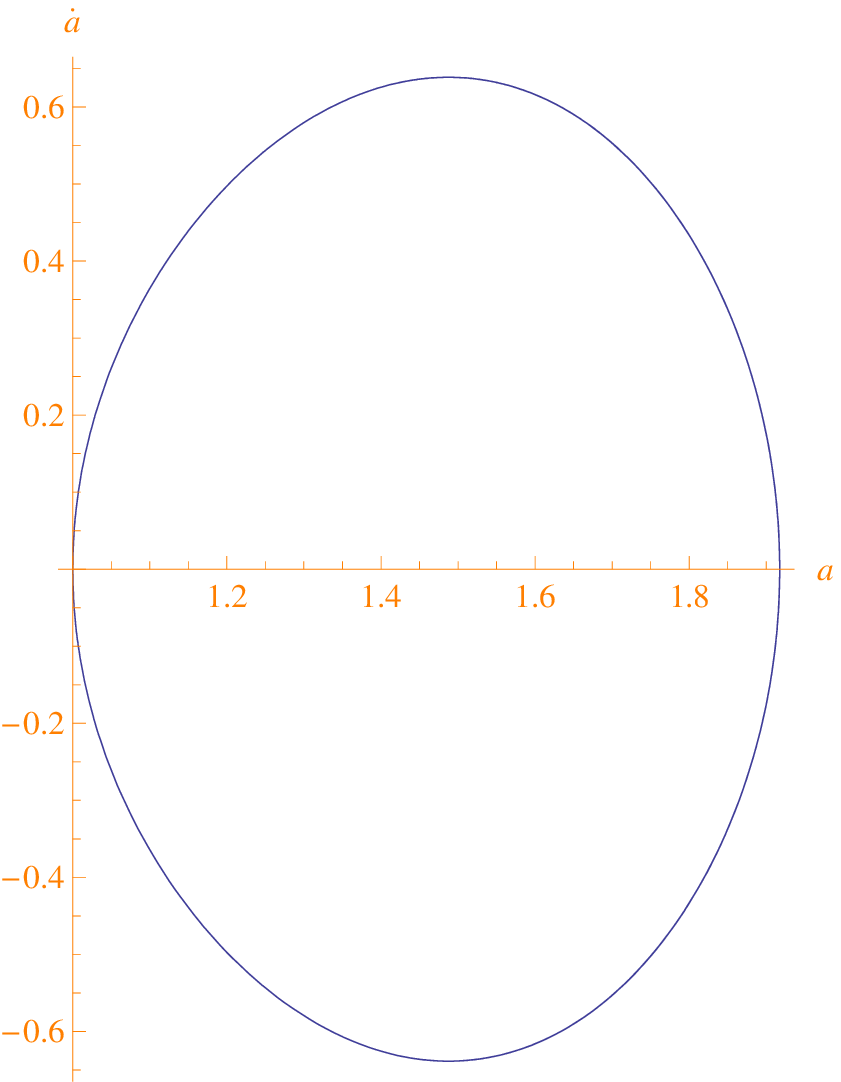}
  \caption{The time evolutionary behavior of the scale factor (left) and the phase space ($a$, $\dot{a}$) diagram (right) for the case $\kappa=1$ and { $\omega=-3.3$} (exotic matter) in minimal massive bigravity model. We have taken the initial values $\dot{a}(0)=0$ and $a(0)=1$, with $M_{Pl}\sim 10^{19}{\rm{Gev}}$ and $m\sim30~ {\rm{Gev}}$.}
  \label{stable1}
\end{figure*}
\textbf{Case}~\textbf{2}~\textbf{:}~\textbf{ $\kappa=-1$}\\

Similarly, using the fixed points in Tables 4 and 5 and the allowed ranges of  $\omega$ and $\bar{\gamma}$, we may repeat the calculation of Case 1 { to obtain the  second Friedmann equation in the case $\kappa=-1$ as follows}
\be\label{Fbi65}
2\frac{\ddot{a}}{a}+\frac{\dot{a}^{2}}{a^{2}}\left(1+3\omega\right)+\frac{1}{a^{2}}\left(1-3\omega\right)+m^{2}\left(-3+4\gamma+
\frac{2\dot{\gamma}}{\frac{\dot{a}}{a}}+3\omega\left(\gamma-1\right)\right)=0.
\ee
In Figs.2, 3, 4, we have also plotted typically the time evolution of the scale factor according to the above equations to show the avoidance of the big bang singularity  for open universe at { stiff matter}, radiation and matter dominant eras,
 respectively. Similar to the previous case,
  the stability of scale factor $a$ indicates for the stability of scale
factor $b$. \\
\begin{figure*}[ht]
  \centering
  \includegraphics[width=2.3in]{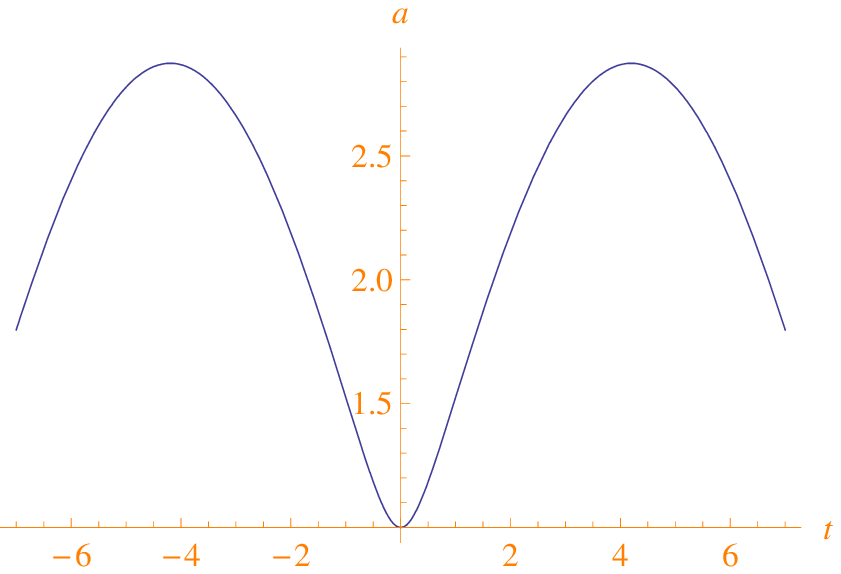}\hspace{2cm}
  \includegraphics[width=2in]{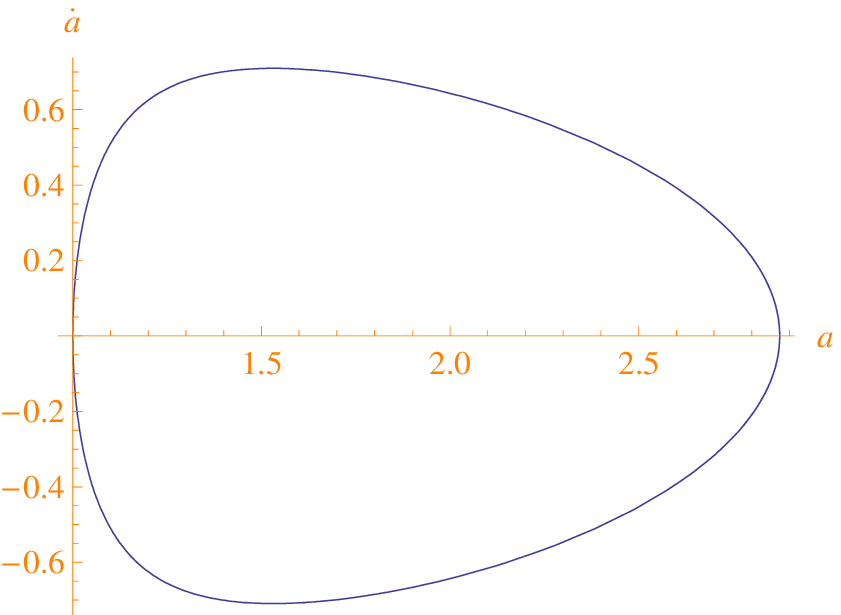}
  \caption{The time evolutionary behavior of the scale factor (left) and the phase space ($a$, $\dot{a}$) diagram (right) for the case $\kappa=-1$ and $\omega=1$ (stiff matter) in minimal massive bigravity model. We have taken the initial values $\dot{a}(0)=0$ and $a(0)=1$, with $M_{Pl}\sim 10^{19}{\rm{Gev}}$ and $m\sim 10~ {\rm{Gev}}$.   }
  \label{stable2}
\end{figure*}

\begin{figure*}[ht]
  \centering
  \includegraphics[width=2.3in]{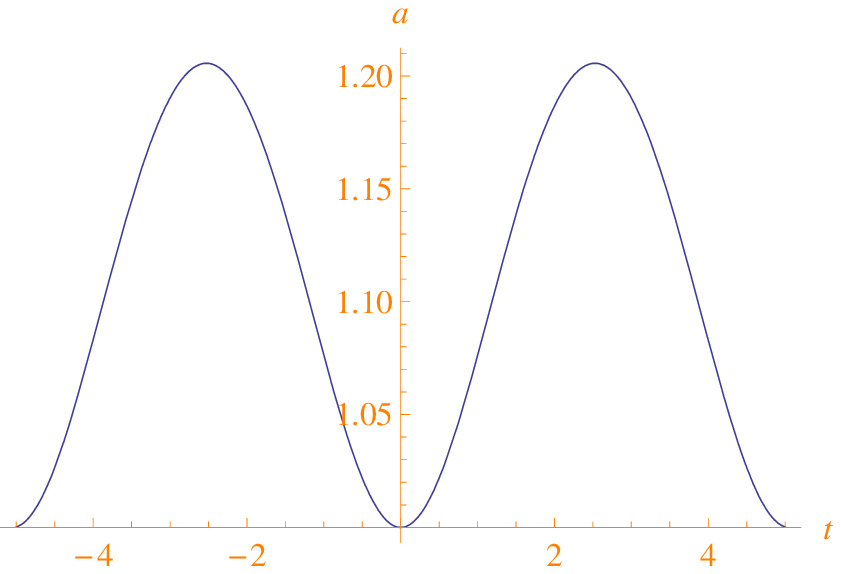}\hspace{2cm}
  \includegraphics[width=1.7in]{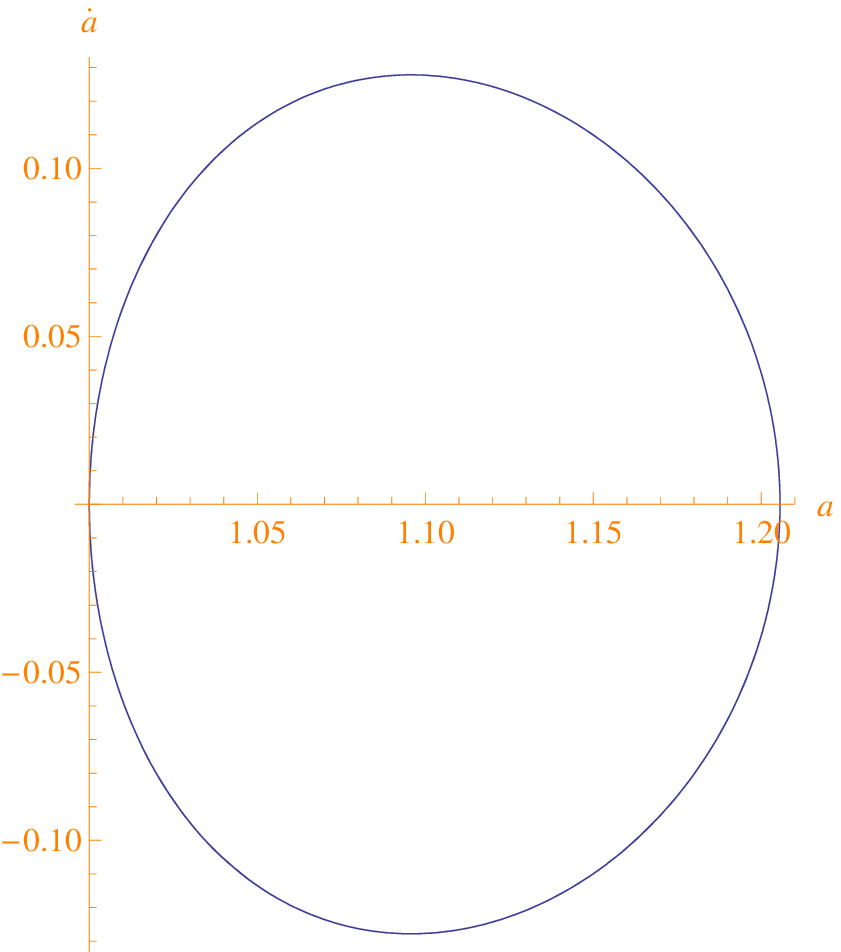}
  \caption{The time evolutionary behavior of the scale factor (left) and the phase space ($a$, $\dot{a}$) diagram (right) for the case $\kappa=-1$ and  $\omega=\frac{1}{3}$ (radiation) in minimal massive bigravity model.
We have taken the initial values $\dot{a}(0)=0$ and $a(0)=1$, with $M_{Pl}\sim 10^{19}{\rm{Gev}}$ and $m\sim 10~ {\rm{Gev}}$.}
  \label{stable3}
\end{figure*}

\begin{figure*}[ht]
  \centering
  \includegraphics[width=2.5in]{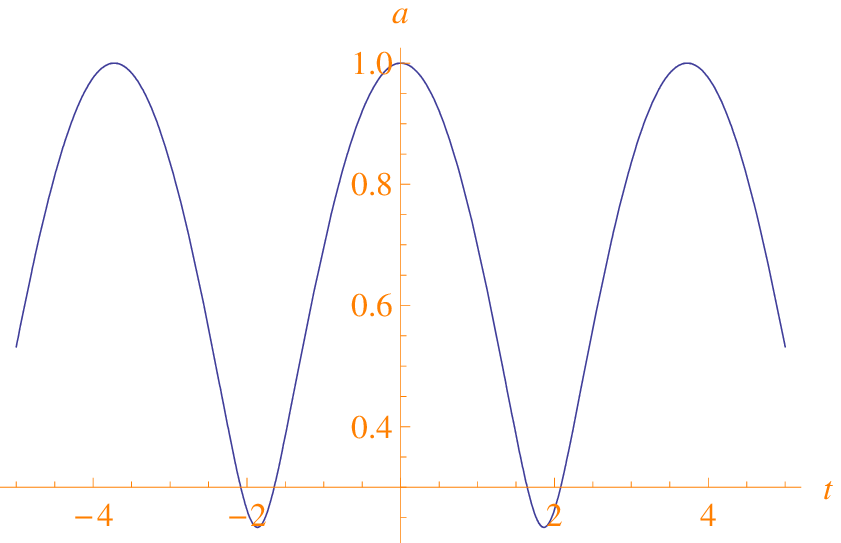}\hspace{2cm}
  \includegraphics[width=1.2in]{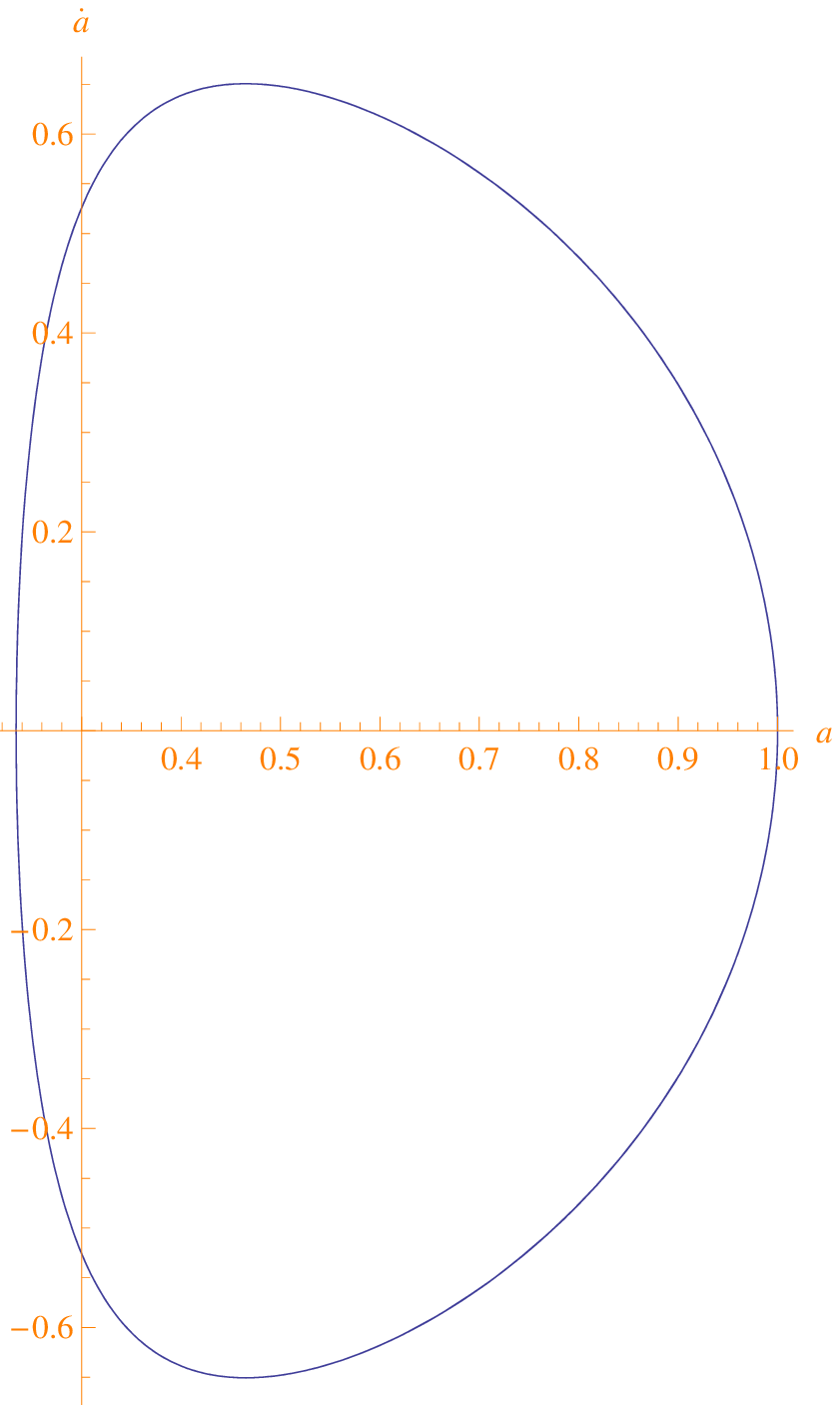}
  \caption{The time evolutionary behavior of the scale factor (left) and the phase space ($a$, $\dot{a}$) diagram (right) for the case $\kappa=-1$ and  $\omega=0$ (dust) in minimal massive bigravity model. We have taken the initial values $\dot{a}(0)=0$ and $a(0)=1$, with $M_{Pl}\sim 10^{19}{\rm{Gev}}$ and $m\sim 10~ {\rm{Gev}}$. }
  \label{stable4}
\end{figure*}

\section{Conclusions\label{Sec6}}
We have studied the cosmological Einstein static solutions of massive bigravity theory. The modified cosmological equations, Friedmann-Robertson-Walker (FRW) universe have been extracted with two scale factors $a$ and $b$. We have found the initial critical scale factor $\bar{a}_{min}$ for the simplest mass term form of massive bigravity in which a large graviton mass is required to have $\bar{a}_{min}$, as small as possible, describing an initial non-singular universe. Moreover, since we have considered a bimetric theory with a couple of metrics and also a couple of modified first and second Friedmann equations, we have used more degrees of freedom and hence the extracted critical scale factor $\bar{a}_{min}$ has shown dependence on the allowed ranges of two quantities, $\omega$ and $\bar{\gamma}$ given in the Table1 and Table2. Moreover, in order to  avoid the highly non-standard matter with $\omega \in(-12,-3)$ in the study of stable solutions for $\kappa=\pm1$, we can select from Table 5 those ranges of $\{\omega$ - $\bar{\gamma}\}$ given by $\{[-0.27,+\infty)$-$(0,0.3]\}$ which is more viable physically because it includes the radiation and ordinary matter for which we have shown the stability in Fig.3 and Fig.4, typically
for the open universe.

Similar to general relativity, in minimal massive bigravity theory we have found that we are not allowed to consider the vanishing curvature of space $\kappa=0$ in our background to obtain the Einstein static universe. Having assumed a perfect fluid with a constant equation of state, beside the assumption that $\bar{\gamma}$ becomes constant including the critical points $\bar{a}$ and $\bar{b}$, we have found the allowed ranges for $\omega$ and $\bar{\gamma}$ to give the stable Einstein static universes for closed
and open universes. Eventually, we have plotted the numerical behavior of $a$ and $\dot{a}$ using the allowed ranges of  $\omega$ and $\bar{\gamma}$ for which we have  Einstein static universe with stability. As is obvious in Fig.1 and Fig.2, for the closed and open Einstein static universes, we have almost similar behaviors for $a$ and $\dot{a}$ in the context of massive bigravity. In both figures the left panels shows the oscillatory evolution of $a$ with respect to time and the right panel shows the trajectory in the ($\dot{a}$,$a$) plane. This numerical study shows properly that massive bigravity can contain stable Einstein static solutions.

The constructed model indicates that the universe can oscillate indefinitely about an
initial static state (fixed point). This result raises the question of finding a
mechanism to terminate the regime of infinite oscillations and start  the  expanding phase which is currently experienced by the universe \cite{Lidsey}. This goal can be achieved by noticing that the two parameters $\omega$ and $\bar{\gamma}$ in equation (\ref{Fbi53}) can vary in such a way that \rm{trM} in (\ref{Fbi52}) becomes non-negative
and so we get  saddle points or repellers, instead of attractors, which may lead to
the  expanding phase.
Such variations may be expected to occur at early universe as a classical phenomena or as a quantum effect. Even, one may propose a mechanism through which
the graviton mass can play the role of a order parameter which is decaying
from a large value to  small values and at a suitable small value, $\rm{trM}$ becomes non-negative and then the attractor solutions change to saddle point or repeller solutions. In either case, it is expected that we get a probability to break the regime of indefinite oscillations and start  the  expanding phase
(inflation),
however the study of such mechanisms are well beyond the scope of this paper.

In order to study  the early universe,
we have assumed the graviton mass to be  large enough to avoid the problem
of low strong coupling scale. To keep the stability issue  to be safely treatable in a classical way at early universe, this large mass should be sufficiently small in comparison to  the Planck mass, otherwise a Planck mass graviton will struggle to have any significant quantum gravitational effect on the background classical evolution.

Finally, it is worth mentioning to the issues about the ghost and gradient instabilities. Based on a purely
background analysis in this paper, it is not possible to study the ghosts or gradient instabilities.   In fact, only finite and infinite branches together with their ghost and gradient instabilities have been studied in bigravity,
whereas all other branches including bouncing cosmologies or a static universe in the asymptotic past or future, so called exotic branches, and their ghost and gradient instabilities have not yet been studied in bigravity. The Einstein static universe is an example of exotic branches for which the ghost and gradient instabilities have not  yet been studied. It is appealing
to confront with such investigation, however a full stability analysis (including
ghost and gradient instabilities) of Einstein static universe in bigravity is well beyond
the scope of this paper and needs a throughout investigation in a future work.

\section*{Acknowledgments}

We would like to thank the anonymous referees whose careful and useful comments led to a very improved revision. We also would like to thank Y. Akrami and N. Khosravi for their valuable, constructive and enlightening comments.


\end{document}